\documentclass[aps,prl,twocolumn,showpacs,superscriptaddress,groupedaddress]{revtex4}  % for review and submission
\usepackage{graphicx}  % needed for figures
\usepackage{dcolumn}   % needed for some tables
\usepackage{bm}        % for math
\usepackage{amssymb}   % for math
\usepackage{amsmath,amsthm,latexsym,amssymb,amsfonts,epsfig}
\usepackage{float}
\usepackage{adjustbox}
\newcommand{\be}{\begin{equation}}
\newcommand{\ee}{\end{equation}}
\newcommand{\bea}{\begin{eqnarray}}
\newcommand{\eea}{\end{eqnarray}}
\begin{document}

% The following information is for internal review, please remove them for submission
\widetext
%\leftline{Version xx as of \today}
%\leftline{Primary authors: Ido Ben-Dayan, Merav Hadad}
%\leftline{To be submitted to PRL}
%\leftline{Comment to {\tt d0-run2eb-nnn@fnal.gov} by xxx, yyy}
%\centerline{\em D\O\ INTERNAL DOCUMENT -- NOT FOR PUBLIC DISTRIBUTION}

% the following line is for submission, including submission to the arXiv!!
%\hspace{5.2in} \mbox{Fermilab-Pub-04/xxx-E}

\title{Thermodynamical Interpretation of the Second Law for Cosmology}
%\input author_list.tex       % D0 authors (remove the first 3 lines
                             % of this file prior to submission, they
                             % contain a time stamp for the authorlist)
                             % (includes institutions and visitors)
\author{Ido Ben-Dayan\footnote{ido.bendayan@gmail.com}},
\affiliation{Physics Department, Ariel University, Ariel 40700, Israel}
\author{Merav Hadad\footnote{meravha@openu.ac.il}}
\affiliation{Department of Natural Sciences, The Open University of Israel, Raanana 43107, Israel}
%\emailAdd{michal.artymowski@fuw.edu.pl}
%\emailAdd{ido.bendayan@gmail.com}}
\date{\today}

\begin{abstract}
The area of a future holographic screen increases monotonically. Associating this area with entropy results in a generalized second law for Cosmology (GSLC). Unlike black hole horizons, screens relevant to Cosmology have no thermodynamical interpretation. We suggest   
relating the entropy of the screens to spacetime degrees of freedom surface density derived from a (seemingly) phase space. This construction enables us to identify the entropy of any holographic screen as the entropy detected by accelerating observers due to their acceleration. Using Unruh's temperature and the equivalence principle, this gives the holographic screens' temperature and yields a thermodynamical interpretation of the GSLC.
\end{abstract}

%\pacs{}
\maketitle
\section{Introduction}

The connection between thermodynamics and gravity dates
back to the seminal paper of Bekenstein showing that black holes have entropy leading to the generalized second law of thermodynamics (GSL) \cite{Bekenstein:1972tm}. 
The relation between the black hole entropy and its area is:
\be
S_{BH} = \frac{A_{EH}}{4G}~,
\end{equation} 
where $\hbar=1$, $S_{BH}$ is the entropy of the black hole, and $A_{EH}$ is its area with its radius being the event horizon.  Incorporating it into a Generalized Second Law of
thermodynamics (GSL) results in,
\be
dS_{gen} \geq 0 ,
\ee
where 
$S_{gen} \equiv S_{out} +
\frac{A_{EH}}{4G} $,
and $S_{out}$ is the von Neumann entropy of the
matter outside the black hole. Hence,
when matter falls into the black hole, an
increase in the horizon area can compensate for the loss of
matter entropy. Therefore, the GSL prevents what would
otherwise be a violation of the (ordinary) second law of thermodynamics to
observers outside the event horizon. Because the entropy of a black hole is proportional to its area rather than its volume, major attempts have been carried out to generalize this notion to other space domains. 

The GSL depends on the event horizon, or some form of a causal horizon such as the Rindler horizon.
These horizons depend on the distant future, which is unsatisfactory in the context of locality. Furthermore, in Cosmology this future is unknown. 
Thus, Cosmology is an immediate example where the GSL and area law cannot be applied. A more local version of the area theorem and the GSL, without using the event horizon, was derived using the construction of holographic screens.
Holographic screens are defined quasi-locally,  and obey an area theorem if the Null Energy 
Condition (NEC) holds \cite{Bousso:2015mqa,Bousso:2015qqa}. Therefore, they also obey a 
GSL for Cosmology (GSLC) \footnote{In \cite{Bousso:2015eda}, the NEC was dispensed and a more general GSLC was derived for Q-screens, that were defined as a "quantum corrected" holographic screens.}.
Contrary to Bekenstein's original GSL that was deduced from thermodynamical considerations, the GSLC does not have a thermodynamical counterpart \cite{Bousso:2018bli}.

Considering the entropy of systems, there is a well-known covariant entropy bound \cite{Bousso:1999xy}, that limits the entropy of null hypersurfaces in arbitrary space-times. The bound could be interpreted as a limit on the number of allowed degrees of freedom, making the question of the phase space of gravity a highly relevant one. Additionally, efforts have tried to express thermodynamical quantities related to gravity in terms of phase space variables and microstates. Such phase space constructions of black holes and accelerated observers have been carried out: For black holes, the constructions are in agreement with the black hole Wald's entropy formula \cite{Brustein:2012sa}, and were used to obtain expected conical singularities in the D1D5 black hole originated form string excitations \cite{Hadad:2015gtr}. For accelerated observers the phase space suggestion matches the degrees of freedom (DoF)  surface density \cite{Hadad:2018rrf}. Both derivations are valid also in generalized theories of gravity. 

In this short note, we propose a thermodynamical description of the GSLC.
We show that a (seemingly) phase space: the extrinsic curvature and its canonical conjugate, can be used in order to construct the relevant entropy expected from the holographic screen. This suggests that the entropy of holographic screens is originated from a phase space which is a first step for establishing thermodynamical interpretation of the area law for Cosmology \footnote{ Let us note, that a canonical transformation between our variables which are the extrinsic curvature and its canonical conjugate, and the more commonly used one which is the induced metric and its canonical conjugate is lacking. However, choosing the extrinsic curvature and its conjugate as phase space does reproduce the Noether charge entropy formula \cite{Hadad:2018rrf}.}. Moreover, since this same phase space can be related to accelerating observers, this construction enables us to identify the entropy of the holographic screens as the entropy detected by accelerating observers due to their acceleration. Thus, having identified the
 	acceleration relevant to each holographic screen, on the one hand, and using
 	Unruh's temperature and the equivalence principle on the other hand we can relate to each holographic screen a local temperature. 
  This leads to the final step:
Using Jacobson's arguments in \cite{Jacobson:1995ab} one finds that the first law of thermodynamics $\delta Q=TdS$ also applies for the entropy defined in \cite{Bousso:2015mqa,Bousso:2015qqa} when $T$ is the holographic screens' temperature. This provides the thermodynamical interpretation of the GSLC.
 
The paper is organized as follows. We start by reviewing the Bousso-Engelhardt construction of screens, the new area law and previous constructions of gravitational phase spaces using the extrinsic curvature and its canonical conjugate. 
We then match between the two and find the conditions that the area law has an interpretation in terms of phase space. 
We give a few examples and then discuss the relevance of our findings to the temperature of the holographic screens. 

\section{A new area law in general relativity}
In \cite{Bousso:2015mqa,Bousso:2015qqa}, Bousso and Engelhardt proved a new area law that is applicable for Cosmology and other situations where the black hole area law fails.
They defined  two kinds of holographic screens: future and past.
Future holographic screens arise in gravitational collapse while
past holographic screens exist in our own expanding universe.

A future holographic screen $H$ is a smooth hypersurface admitting a foliation by marginally trapped surfaces called leaves and
a past holographic screen by marginally anti-trapped surfaces. 
A marginally trapped surface is a co-dimension 2 compact spatial
surface $\sigma$ whose two future directed orthogonal null geodesic congruences satisfy 
\bea
\theta_{k}\equiv\hat{\nabla}_{a}k^{a} =0;\quad
\theta_{l}\equiv \hat{\nabla}_{a}l^{a}<0,
\eea
where $k^{a}$ and $l^{a}$ are the two future directed null vector fields orthogonal
to $\sigma$, $\theta_{k}$ and $\theta_{l}$
are the null expansions, and $\hat{\nabla}_{a}$ is computed
with respect to the induced metric on $\sigma$. 
They also defined a tangent vector field $h^{a}$ on $H$ which is written as a (unique) linear
combination of the two null vector fields orthogonal to each leaf:
$h^{a}=\alpha l^{a}+\beta k^{a}$ and fixed the normalization of $h^a$ by requiring
that the function $r$ increases at unit rate along $h^{a}$, $h(r)=h^{a}(dr)_{a}=1$. 
The leaves are labeled by $\sigma(r)$. 
In this way they get a (non-unique) evolution parameter
$r$ along the screen $H$ such that $r$ is constant on any leaf
and increases monotonically along the fibers $\gamma$, (a fibration of $H$).
They then proved the area law: 
\emph{The area $A$ of the leaves of any regular future
holographic screen $H$ increases strictly monotonically}.

Moreover, the construction implies more specifically, that
the area of leaves increases at the rate
\be
\frac{dA}{dr}=\int_{\sigma(r)}\sqrt{h^{\sigma(r)}}\alpha\theta_{l}^{\sigma(r)}>0,
\ee
%where $h_{ab}^{\sigma(r)}$ is the induced metric on the leaf $\sigma(r)$
%and $h^{\sigma(r)}$ is its determinant.
where $h^{\sigma(r)}$ is the determinant of the induced metric on the leaf $\sigma(r)$ \footnote{Past holographic screens also obey an appropriate area theorem and our thermodynamical interpretation discussed here is valid for them as well.}.

\section{The entropy surface density as a gravitational phase space} 

In different aspects of gravity the gravitational entropy surface density can be regarded as a gravitational phase space. To start
with, this phase space can be related to the surface density of space
time degrees of freedom (DoF) which are expected to be observed by
an accelerating observer in curved spacetime \cite{Padmanabhan:2010xh}.
This DoF surface density was first derived by Padmanabhan for a static
spacetime using thermodynamical considerations. It was found 
that, if the foliation of spacetime is done with respect to the direction
of the acceleration, then this density can also be constructed from
a specific extrinsic curvature and its canonical conjugate \cite{Hadad:2018rrf}. %\IB{REMOVE The same
%phase space derivation can also be used in order to obtain black holes entropy \cite{Brustein:2012sa}.
%Moreover, this unique construction was found to be useful even for
%string theory discussions \cite{Mathur:2005zp}. It was also found that some singularities obtained
%due to string theory excitations of a $D1D5$ black hole 
%can also be explained by the uncertainty principle expected from
%the same seemingly phase space \cite{Hadad:2015gtr}. 
%}
Other examples are also discussed in \cite{Hadad:2018rrf,Brustein:2012sa,Hadad:2015gtr,Mathur:2005zp}.

All these example have two things in common: Deriving a phase space
by foliating space-time along a spatial direction (the radial direction
for a black hole and the acceleration direction for accelerating observers),
and identifying the relevant gravitational phase space as the extrinsic
curvature of a specific hypersurface and its canonical conjugate. 

Let us summarize the derivation which relates the
surface density of space time DoF  
%observed by an accelerating observer in curved spacetime 
to the gravitational
phase space: extrinsic curvature and its canonical conjugate.

\subsection{ Gravitational phase space: extrinsic curvature and
its canonical conjugate}

One starts by defining the direction of the space-like vector field in a stationary D-dimensional spacetime. (This direction 
is the acceleration direction for accelerating observers and is
the radial direction of a black hole). In general one
considers a $D$ velocity unit vector field $u^{a}$ and acceleration
$a^{a}=u^{b}\nabla_{b}u^{a}\equiv an^{a}$ (where $n^{a}$ is a unit
vector and $u^{a}n_{a}=0$) \footnote{We assume that both unit vectors: $n^{a}$and $u^{a}$ are hyper surface
orthogonal and thus fulfill Frobenius's theorem.}. 
%One considers $D-2$ hyper-surfaces $\Sigma_{D-2}$ that are also orthogonal to $u_a$ and $n_a$.
One foliates spacetime with respect to the unit vector field
$n_{a}$ by defining a $(D-1)$- hyper-surface $\Sigma_{D-1}$,  which is normal to
$n_{a}$. 
%Next, one reduces another dimension by considering $D-2$ hyper-surfaces $\Sigma_{D-2}$ that are also orthogonal to $u_a$. 
%The lapse function $M$ and shift vector $W_{a}$ satisfy
%$r_{a}=Mn_{a}+W_{a}$ where $r^{a}\nabla_{a}r=1$ and $r$ is constant
%on $\Sigma_{D-1}$. The $\Sigma_{D-1}$ hyper-surfaces metric $h_{ab}$
%is given by $g_{ab}=h_{ab}+n_{a}n_{b}$. 
%The extrinsic curvature of
%the hyper-surfaces is given by %$K_{ab}=-\frac{1}{2}\mathcal{L}_{n}h_{ab}$
%where $\mathcal{L}_{n}$ is the Lie derivative along $n^{a}$. We then reduce another dimension by considering $D-2$ hyper-surfaces $\Sigma_{D-2}$ that are also orthogonal to $u_a$. 
%The $\Sigma_{D-2}$ metric $\sigma_{ab}$ is given by $\sigma_{ab}=h_{ab}+u_{a}u_{b}$. 

As was first noted by Brown \cite{Brown:1995su} for generalized theories
of gravity, the canonical conjugate variable of the extrinsic curvature
$K_{bc}$ is $4\sqrt{-h}n_{a}n_{d}U_{0}^{abcd}$.  $U_{0}^{abcd}$
is an auxiliary variable, which equals $\frac{\partial\mathcal{L}}{\partial R_{abcd}}$
when the equations of motion hold. From \cite{Hadad:2018rrf}
the relevant phase space for detectors with D-velocity $u^{a}$ at
point $P$ can be identified by projection of the extrinsic curvature
tensor and its canonical conjugate variable on the vector field $u^{a}$\footnote{\label{18}This means that we distinguish these canonically conjugate variables
from the others by projecting the extrinsic curvature and its canonical
conjugate variable along the time-like unit vector $u_{b}$. Actually
this should be done more carefully since the Lie derivative of the
normal vector $u_{b}$ does not vanish in general and thus leads,
for example, to a contribution to the canonical conjugation of the induced metric $h_{ab}$.}:

\begin{eqnarray}
\left\{ K^{nm}u_{m}u_{n},4\sqrt{h}U_{0}^{abcd}n_{a}u_{b}u_{c}n_{d}\right\}. %.\label{canonicalstring}
\label{tensor canonical tern-1}
\end{eqnarray}
The gravitational degrees of freedom density detected by an accelerating
detector with D-velocity $u^{a}$ at point $P$ is constructed from
multiplying these special canonically conjugate variables.
Thus, using $K^{ab}u_{b}u_{a}=n^{a}a_{a}=a$, the gravitational $D-2$
surface density of the spacetime DoF observed by an accelerating observer
$\Delta n$ per unit time $\Delta t$ is
\begin{eqnarray}
\frac{\Delta n}{\Delta t} & = & 4a\sqrt{h}U_{0}^{abcd}n_{a}u_{b}u_{c}n_{d}, \label{density per time-1}
\end{eqnarray}
where the $D-2$ hyper-surface is orthogonal to both $u_a$ and $n_a$.
Finally, using the Euclidean limit and integrating over Euclidean time, the expected spacetime $D-2$ hyper-surfaces entropy density for accelerating observer was derived \cite{Hadad:2018rrf}. This proves that this entropy is
constructed from the extrinsic curvature and its canonical conjugate as long as they
are derived by foliating spacetime with respect to the direction of
the acceleration. 

\section{The area law and the gravitational phase space} \label{sec:phasespace}

We have seen that for certain $D-2$ hyper-surfaces, one can construct a phase space using the extrinsic curvature and its canonical conjugate,
while in \cite{Bousso:2015mqa,Bousso:2015qqa} $D-2$ hypersurfaces were used as leaves to prove the area law. We therefore wish to find what are the conditions where these hypersurfaces are the same. If so, then we have succeeded in constructing the phase space associated with the area growth.

We start with the vector $h^{a}$ (defined in  \cite{Bousso:2015mqa,Bousso:2015qqa} as $h^{a}=\alpha l^{a}+\beta k^{a}$ where $k^{a}$ and $l^{a}$
are the two future directed null vector fields orthogonal to $\sigma$) and rewrite it in terms of a non-null unit vector $u_{a}$ as  $h_{a}=Nu_{a}+V_{a}$ where $u_{a}$ is a vector field orthogonal to $\sigma$ (and thus $V_{a}$ is also normal to $\sigma$). We choose the direction of $u_{a}$ so that the direction of its acceleration, namely $a_{b}=u^{a}\nabla_{a}u_{b}$, is a vector field orthogonal to $\sigma$ and to $u_{a}$. The magnitude of the acceleration is given by $a=\sqrt{a^{b}a_{b}}$  and we define its direction by the unit vector $n_{a}=a_{a}/a$.
 One can always find such unit vectors $n_{a}$ and $u_{a}$ which are normal to each other, to $\sigma$, and fulfills $u^{a}\nabla_{a}u_{b}=an_{b}$.

Next we construct our foliation using the two unit vector fields $n_{a}$ and $u_{a}$. We start by foliating spacetime with respect to the unit vector field $n_{a}$.
In order to do that we define a $\Sigma_{D-1}$ hyper-surfaces. The $\Sigma_{D-1}$ hyper-surfaces metric $h_{ab}$ is given by $g_{ab}=h_{ab}+n_{a}n_{b}$.
  Its lapse function $M$ and shift vector $W_{a}$ satisfy $t_{a}=Mn_{a}+W_{a}$
where $t^{a}\nabla_{a}t=1$ and $t$ is constant on $\Sigma_{D-1}$.
The extrinsic curvature of the hyper-surfaces is given
by $K_{ab}=-\frac{1}{2}\mathcal{L}_{n}h_{ab}$ where $\mathcal{L}_{n}$
is the Lie derivative along $n^{a}$. The $\Sigma_{D-2}(\equiv\sigma)$
hyper-surfaces metric $\sigma_{ab}$ is given by $h_{ab}=\sigma_{ab}-u_{a}u_{b}$.
The lapse function $N$ and shift vector $V_{a}$ satisfy $h_{a}=Nu_{a}+V_{a}$
where $h^{a}D_{a}r=1$ and $r$ ( and also $t$) are constant on $\Sigma_{D-2}$
and $D_{a}=h_{ab}\nabla^{b}$ is the derivative computed with respect
to the induced metric on $\Sigma_{D-1}$. Note that since the vector $h^{a}$ also satisfies  $h^{a}=\alpha l^{a}+\beta k^{a}$ where $k^{a}$ and $l^{a}$
are the two future directed null vector fields orthogonal to $\sigma$, we find that the shift vector $V_{a}$ may only have a component along $n_{a}$ and thus we may write  $V_{a}=Vn_{a}$.

To summarize, our induced $D-2$ metric is defined as:
\be
\sigma_{ab}=g_{ab}+\left(u_{a}u_{b}-n_{a}n_{b}\right). \label{eq:sigma}
\ee 
On the other hand, the $D-2$ metric discussed in \cite{Bousso:2015qqa} should be orthogonal to both $l_a$ and $k_a$. For $l_ak^a=-1$, a natural candidate is:
\be
q_{ab}=g_{ab}+\left(l_ak_b+k_al_b\right). %\Rightarrow C=-\frac{1}{l_ak^a}
\ee
One can always find $\tilde{\alpha}$ and $\tilde{\beta}$ which relate the null vectors  $l_a$ and $k_a$ to the unit vectors  $u_a$ and $n_a$:
\be
u^{a}=\tilde{\alpha} l^{a}+\tilde{\beta} k^{a}; \label{eq:defin u}\quad
n^{a}=\tilde{\alpha} l^{a}-\tilde{\beta} k^{a} %\label{eq:defin n}
\ee
Requiring $\sigma_{ab}=q_{ab}$ only imposes a normalization condition:
\be
2\tilde{\alpha} \tilde{\beta}=1. \label{eq:C} 
\ee
Note that this also leads to:
\be
\theta_n=\theta_u=\tilde \alpha \theta_l. \label{eq:smalltheta}
\ee

Similar to the analysis of black holes and accelerated observers, we suggest that the relevant entropy density related to trajectories
along the unit vector field $u^{a}$ can be constructed by these conjugate
variables $K_{bc}$ and $4\sqrt{-h}n_{a}n_{d}U_{0}^{abcd}$, at point $P$ after projecting them along $u^{a}$ \footnote{The admonition of the previous comment holds here as well.}:  
\be
\left\{ K^{nm}u_{m}u_{n}(x),4\sqrt{h}U_{0}^{abcd}n_{a}u_{b}u_{c}n_{d}(x)\right\}, %.\label{canonicalstring}
\label{tensor canonical term}
\ee
where we mark the coordinates by $(t,r,x)=(t,r,x_{1},...,x_{D-2})$.
Since $K^{ab}u_{b}u_{a}=-u_au^b\nabla_b n^a=a$ where by construction $a_{a}=u_{b}\nabla^{b}u_{a}=an_a$ we deduce  
that the gravitational density degrees of freedom along the direction
$h^{a}$ at point $P$ is constructed from multiplying these special
canonically conjugate variables.
Thus, the gravitational $D-2$ surface density of the spacetime DoF, $\Delta n$,
obtained due to varying along the direction $h_{a}=Nu_{a}+V_{a}$
per unit ``time''
$r$ is
\begin{eqnarray}
\frac{\Delta n}{\Delta r} & = & 4aN\sqrt{\sigma}U_{0}^{abcd}n_{a}u_{b}u_{c}n_{d}(x),\label{density per time}
\end{eqnarray}
where $\sqrt{h}=N$$\sqrt{\sigma}$, since $N$ is the lapse function
of the direction of the vector $u^{a}$.
For Einstein theory where $\mathcal{L}=\frac{1}{16\pi G}R$, using
$U_{0}^{abcd}=\frac{\partial\mathcal{L}}{\partial R_{abcd}}=\frac{1}{16\pi G}\frac{1}{2}\left(g^{ac}g^{bd}-g^{ad}g^{bc}\right)$,
we find that $U_{0}^{abcd}n_{a}u_{b}u_{c}n_{d}%=\frac{1}{32\pi G}\left(g^{ac}g^{bd}-g^{ad}g^{bc}\right)n_{a}u_{b}u_{c}n_{d}
=\frac{1}{32\pi G}$,
and thus
\be
\frac{\Delta N_0}{\Delta r}=\int_{\sigma(r)}\frac{\Delta n}{\Delta r}=\int_{\sigma(r)}\frac{1}{8\pi G}aN\sqrt{\sigma}(x),
\label{eq:entropy_growth}
\ee
where $N_0$ is the number of DoF on the area of the screen.

On the other hand, according to \cite{Bousso:2015mqa,Bousso:2015qqa} the area growth of the holographic screens is given by 
\be
\frac{\Delta A}{\Delta r}=\int_{\sigma(r)}\alpha\theta_{l}\sqrt{\sigma}(x). \label{eq:area_growth}
\ee
Note that it is expected that in order to obtain from \eqref{eq:area_growth} the rate of entropy growth one should divide it by $4G$. 
 
 Finally, we demand that  the rate of change of
 the entropy of the holographic screens will be the same as the rate of change of their gravitational DoF along the same direction, and equate \eqref{eq:entropy_growth} to \eqref{eq:area_growth} divided by $4G$. We find:
\be
a=\eta N^{-1}\alpha\theta_{l} \label{eq:thetahna}
\ee
where we introduce a constant of proportionality $\eta$ between the entropy and the gravitational DoF. From now on, we set $\eta=1/4$ because it will reproduce the Schwarzschild black hole temperature for $\alpha=-1$. 

Equation \eqref{eq:thetahna} is the major result of this work. It provides an algorithm that associates the area growth of holographic screens to the density growth of the gravitational phase space observed by accelerated observers. Moreover, as we will see, this allows us to prove that the entropy of the holographic screens can be interpreted as the entropy of accelerated observers and thus provide the desired thermodynamical interpretation. 

It is easy to prove that for any given $\alpha$ one can find a $\beta$ that will give a direction of an acceleration  $n^a$ (i.e. to give a physical (i.e. positive) solution to $\tilde \alpha ^2$ and  $\tilde \beta ^2$). To see this note that 
 $u_au_b=\sigma_{ab}-g_{ab}+n_{a}n_{b}$,  and thus $a=\Theta_n-\theta_n$. Using \eqref{eq:smalltheta} we arrive at the following equation 
 
\bea
\Theta_n=\left(\tilde \alpha+ N^{-1}\alpha/4\right)\theta_l
\eea
using the customary notation of the expansion rate   $\Theta_x\equiv\nabla_ax^a$.
Moreover, since  $V^a$, the shift vector of $h^a$ is orthogonal to $u^a$, we can extract the lapse function, $N$ as a function of $\alpha, \tilde \alpha,\beta, \tilde \beta$ via:
$N=-h_au^a=-(\alpha l_au^a+\beta k_au^a)=\alpha \tilde \beta+\beta \tilde \alpha$ yielding
\be
\Theta_n=\left(\tilde \alpha+\left(\alpha \tilde \beta+\beta \tilde \alpha\right)^{-1}\alpha/4\right)\theta_l. \label{eq:major}
\ee
 For constant $\tilde \alpha$ and $\tilde \beta$, one finds $\Theta_n=\tilde \alpha \Theta_l-\tilde \beta \Theta_k$. Using the normalization condition  \eqref{eq:C}, we find that for $\Theta_l\neq\theta_l$:
\bea
2\tilde \alpha ^2_{1,2}&=&\frac{\frac{\alpha}{\beta} \left( \Theta_l-3/2\theta_l\right)-\Theta_k}{2\left(\Theta_l-\theta_l\right)}
\nonumber\\
&\pm& \frac{\sqrt{ \left( \frac{\alpha}{\beta} \left( \Theta_l-3/2\theta_l\right)-\Theta_k \right)^2+4 \frac{\alpha}{\beta}
		\Theta_k\left(\Theta_l-\theta_l\right)}}{2\left(\Theta_l-\theta_l\right)}\cr\label{eq:majorsimple}
\eea
while for $\Theta_l=\theta_l$:
\be
2\tilde \alpha ^2=\frac{-\alpha \Theta_k}{\alpha/2 \Theta_l+\beta\Theta_k}.\label{eq:majorsimple1}
\ee
Obviously, for any given $\alpha$ one can always find a family of $\beta$-s that gives a physical (i.e. positive) solution to $\tilde \alpha ^2$. This proves that for any holographic screen one can relate a family of accelerating observers (that accelerate along $n^a$). These accelerating observers will relate  (the growth) of the holographic screens to the (growing of the) expected entropy due to their acceleration.  

\section{Examples} \label{sec:examples}
Let us now demonstrate this construction in a few examples and explicitly construct the different vectors. 
In each example, we define two null vectors  $l_a$ and $k_a$ and calculate the relevant expansions rates. Next, by using the conditions in  \cite{Bousso:2015mqa,Bousso:2015qqa} we find the holographic screen relevant to the null vectors. Finally,  we use \eqref{eq:majorsimple} or \eqref{eq:majorsimple1} in order to calculate $\tilde{\alpha}$. Note that since \eqref{eq:defin u} and \eqref{eq:C} give
\bea
u^{a}=\tilde{\alpha} l^{a}+(2\tilde{\alpha})^{-1} k^{a} \label{eq:defin u1}
\eea 
this determines the velocity vector field of the accelerated observers.  

%\subsection{Black Hole/Star case} 
\textbf{The Black Hole/Star Example:}
Consider a kind of a black hole in Eddington-Finkelstein coordinates:
\be
ds^2=-f(r)dv^2+2 dv dr+r^2d\Omega,
\ee
where for the Schwarzschild black hole $f(r)=(1-2M/r)$. 
Constructing the two null vectors:
 \be
 l^a=\frac{1}{\sqrt{2}}(0,1,0,0) ;\,
 k^a=\frac{1}{\sqrt{2}}(-2,-f(r),0,0).
 \ee
Calculation of the expansion rates reveals as expected:
\be
\theta_k=-\frac{2f(r)}{r};\,
\Theta_k=-\frac{2f(r)}{r}-f'(r);\,
\Theta_l=\theta_l=-\frac{2}{r}.
\ee
So $\theta_l<0$ always, and $f(r_0)=0$ is the only hypersurface at which $\theta_k=0$. Note that this is the horizon $r_0=2m$ in the Schwarzschild case. Using  \eqref{eq:majorsimple1} we find
\be
2\tilde \alpha^2=\frac{-\alpha f'(r_0)}{\alpha/r_0+\beta f'(r_0)}.
\ee
For the Schwarzschild black hole, this simplifies to 
\be
2 \tilde \alpha^2=\frac{-\alpha}{\alpha+\beta}.
\ee
Since $\alpha<0$, this requires the denominator to be positive, and weakly restricts $\beta$.
Interestingly enough, the above result for $\tilde \alpha$ is valid also for a non-stationary metric such as the Vaidya metric, that describes a  "star" or a "black hole" with infalling or outgoing null shells of energy. The interesting difference is that now the horizon is a time-dependent shell according to $r=2M(v)$ where $v$ is the time-like coordinate.

%\subsection{The cosmological case}
\textbf{The Cosmological Example:}
Consider the FLRW metric,
\[
ds^{2}=-dt^{2}+a^{2}(t)dr^{2}+a^{2}(t)r^{2}d\varOmega^{2}
\]
The null vectors are: 
\[
k^{a}=\frac{1}{\sqrt{2}}(-1,a^{-1}(t),0,0);\,
l^{a}=\frac{1}{\sqrt{2}}(-1,-a^{-1}(t),0,0).
\]
Calculating $\theta_{k},\theta_{l}$ gives 
\be
\theta_{k}=\frac{2-2r\dot{a}}{ra};\quad
\theta_{l}=-\frac{2+2r\dot{a}}{ra}.
\ee
$\theta_{k}=0$ imposes $\dot{a}=1/r$. Hence, $\theta_{l}=-4\frac{\dot{a}}{a}=-4H<0$
(for $H>0$), 
$\varTheta_{k}=-H$
and $\varTheta_{l}=-5H$.
Using \eqref{eq:majorsimple}, since  $\Theta_l\neq\theta_l$ results in:
\be
2\tilde \alpha^2_{1,2}=\frac{-(\alpha+\beta) \pm \sqrt{(\alpha+\beta)^2+4\alpha \beta}}{2\beta}.
\ee

\section{Discussion}

In this paper we have established  
the connection between thermodynamics and holographic screens in the cosmological case. We  suggested identifying the (growing of the) area of the holographic screens to the (growing of the) extra DoF detected by accelerating observers. These extra DoF are encoded in a unique kind of gravitational phase space, which was found to be useful for the entropy of stationary black holes, as well as for accelerating observers in a stationary metric. This gravitational phase space has the advantage that it can be constructed in any spacetime and it is relevant also for the cosmological case and more generally non stationary cases. The fact that we have established a connection between the area and the gravitational phase space proves that the entropy is indeed proportional to the area. 

Using this identification we have found that any holographic screen can be related to a family of accelerating observers. Though the expansion of the holographic screens is unique, the acceleration direction of the accelerating observers is not unique. This can be seen from eq.  \eqref{eq:area_growth}, where for each $\alpha$ which determines the rate of expansion one can find a set of $\tilde{\alpha}$ which depends not only on $\alpha$ but also on $\beta$. Since $\beta$ is (almost) a free parameter in the derivation of the holographic screens, $\tilde{\alpha}$ which determines the direction of the acceleration, becomes also (almost) a free parameter. Note that although our derivation limits $\beta$, in such a way that the square in \eqref{eq:majorsimple},\eqref{eq:majorsimple1} must be positive, we are still left with enough freedom for the direction of the acceleration. Hence, we have managed to relate each holographic screen to a family of accelerating observers and their relevant gravitational phase space.

Having identified the relevant gravitational phase space for the holographic screens in the cosmological case, the next step for constructing their thermodynamical properties is identifying their temperature and verifying its entropy. 
We start with identifying the holographic screens' temperature. Having identified the acceleration relevant to each holographic screen in \eqref{eq:thetahna}, on the one hand, and using Unruh's temperature: $T=Na/2\pi$ and the equivalence principle on the other hand, the most natural identification to the screens' temperature is:  
\be
T=\alpha\theta_{l}/8\pi \label{eq:temperature}
\ee 
where we have used $\eta=1/4$ to match the known result of black hole temperature for $\alpha=-1$.
Note that although our derivation suggests a family of accelerating observers to each holographic screen (which depend on $\alpha$ and $\beta$), the holographic screens construction leads to a specific temperature (which depends only on  $\alpha$). 
 As expected, and can be seen from the Vaidya and cosmological examples, this leads to a time-dependent temperature. 
Finally, having the holographic screens' temperature, we can use Einstein's equations   
and derive the first law of thermodynamics $\delta Q=T\delta S$ as in \cite{Jacobson:1995ab}. As a result, the holographic screens have a well-defined phase space density, entropy and temperature. Hence the GSLC is fully specified by thermodynamical quantities, and we have a thermodynamical interpretation of the GSLC.

\end{document}